\documentclass[epj]{webofc}
\usepackage[utf8]{inputenc}
\usepackage[varg]{txfonts}   
\usepackage{booktabs}
\usepackage{xcolor}
\definecolor{darkred}{rgb}{0.4,0.0,0.0}
\definecolor{darkgreen}{rgb}{0.0,0.4,0.0}
\definecolor{darkblue}{rgb}{0.0,0.0,0.4}

\usepackage{amssymb}
\usepackage{amsmath}
\graphicspath{ {./plots/}  }

\usepackage{etoolbox}
\newtoggle{arxiv}
\toggletrue{arxiv}

\usepackage{xcolor}

\usepackage[bookmarks,linktocpage,colorlinks,
    linkcolor = darkred,
    urlcolor  = darkblue,
    citecolor = darkgreen]{hyperref}

\newcommand{\ifarxiv}[1]{\iftoggle{arxiv}{#1}{}}

\newcommand{\comment}[1]{}

\newcommand{\lr}[1]{ \left( #1 \right) }
\newcommand{\lrs}[1]{ \left[ #1 \right] }
\newcommand{\lrc}[1]{ \left\{ #1 \right\} }
\newcommand{\vev}[1]{ \langle \, #1 \, \rangle }
\newcommand{\cev}[1]{ \langle \langle \, #1 \, \rangle \rangle }

\newcommand{\tr}{ {\rm Tr} \, }

\newcommand{\ket}[1]{ \, | #1 \rangle }
\newcommand{\bra}[1]{ \langle #1 | \, }

\newcommand{\expa}[1]{ \exp{\left( #1 \right)} }

\wocname{EPJ Web of Conferences}
\woctitle{Lattice2017}
\begin{document}
%

\selectlanguage{english}
\title{Real-time dynamics of matrix quantum mechanics beyond the classical approximation}
\author{%
\firstname{Pavel} \lastname{Buividovich}\inst{1}\fnsep\thanks{Speaker. \email{pavel.buividovich@physik.uni-regensburg.de}. This work was supported by the S.~Kowalevskaja award from the A.~{von Humboldt} foundation.}
\and
\firstname{Masanori} \lastname{Hanada}\inst{2,3,4,5}
\and
\firstname{Andreas} \lastname{Sch\"{a}fer}\inst{1}
}

\institute{%
Institute for Theoretical Physics, Regensburg University, D-93053 Regensburg, Germany
\and
Yukawa Institute for Theoretical Physics, Kyoto University,
Kitashirakawa Oiwakecho, Sakyo-ku, Kyoto 606-8502, Japan
\and
Stanford Institute for Theoretical Physics, Stanford University, Stanford, CA 94305, USA
\and
Nuclear and Chemical Sciences Division, Lawrence Livermore
National Laboratory, Livermore, California 94550, USA
\and
The Hakubi Center for Advanced Research, Kyoto University,
Yoshida Ushinomiyacho, Sakyo-ku, Kyoto 606-8501, Japan
}
\abstract{%
 We describe a numerical method which allows us to go beyond the classical approximation for the real-time dynamics of many-body systems by approximating the many-body Wigner function by the most general Gaussian function with time-dependent mean and dispersion. On a simple example of a classically chaotic system with two degrees of freedom we demonstrate that this Gaussian state approximation is accurate for significantly smaller field strengths and longer times than the classical one. Applying this approximation to matrix quantum mechanics, we demonstrate that the quantum Lyapunov exponents are in general smaller than their classical counterparts, and even seem to vanish below some temperature. This behavior resembles the finite-temperature phase transition which was found for this system in Monte-Carlo simulations, and ensures that the system does not violate the Maldacena-Shenker-Stanford bound $\lambda_L < 2 \pi T$, while the classical dynamics inevitably breaks the bound.
}
\maketitle

\section{Introduction}
\label{sec:intro}

Thermalization of strongly interacting quantum systems is one of the important problems in different areas of modern physics, ranging from apparent thermalization of the quark-gluon plasma \cite{Nastase:hep-th/0501068,Schaefer:1012.4753,Kurkela:11:1,Romatschke:1307.2539,Chesler:1309.1439,Kurkela:15:1} in heavy-ion collisions to inflation of our Universe and dynamics of ultra-cold quantum gases \cite{Berges:Science341}. Unfortunately, the number of tools which can be used to study non-equilibrium real-time dynamics of generic non-perturbative, non-supersymmetric and non-integrable quantum field theories (in particular, non-Abelian gauge theories) is rather limited. On the one hand, in the regime of large field strengths (large quantum occupation numbers) and small coupling constants one can reliably use the classical equations of motion \cite{Gelis:11:1,Schafer:09:1}, re-summing secular divergences by averaging over quantum fluctuations in initial conditions \cite{Gelis:11:1}. On the other hand, in the dilute plasma regime with small quantum occupation numbers one can use the kinetic theory description \cite{Moore:hep-ph/0209353}. The intermediate regime between the two descriptions with occupation numbers of order of one is very important for matching early-time strongly non-equilibrium evolution with late-time hydrodynamic behavior \cite{Kurkela:11:1,Kurkela:15:1}.

In these Proceedings, we aim to go deeper into this intermediate regime of neither large nor small occupation numbers. We outline an extension of the classical dynamics approximation which incorporates sub-leading effects with respect to large occupation numbers by approximately taking into account the quantum dispersion of the wave function of the system. In a few words, the basic idea is to approximate the time-dependent Wigner function by the most general Gaussian function with time-dependent mean and dispersion. This approximation is also used to study short-scale real-time processes in quantum chemistry \cite{Broeckhove:THEOCHEM199}. However, to our knowledge, it has not been previously used in the context of high-energy physics. \ifarxiv{For fermionic fields interacting with classical gauge fields, this Gaussian approximation is equivalent to the classical-statistical field theory approximation which is by now a standard tool to study real-time dynamics of fermions \cite{Berges:14:1}.}

As the simplest prototypical system which features thermalization and (at least classical) chaos and still has non-Abelian structure similar to that of Yang-Mills fields \cite{Savvidy:NPB84}, in these Proceedings we consider the matrix mechanics with the Hamiltonian
\begin{eqnarray}
\label{matrix_mechanics_hamiltonian}
 H = \frac{1}{2} \, \sum_i \tr{P_i^2} - \frac{1}{4} \, \sum\limits_{i,j} \tr \lrs{X_i, X_j }^2 ,
\end{eqnarray}
where $X_i$ and $P_i$ are the canonically conjugate coordinates and momenta which take values in the Lie algebra of $SU\lr{N}$ group\ifarxiv{{ }(traceless Hermitian matrices) and are additionally labelled by the coordinate indices $i, j = 1 \ldots d$. Correspondingly, the commutator and the trace operations in (\ref{matrix_mechanics_hamiltonian}) are applied to the matrix indices of $X_i$ and $P_i$}. This Hamiltonian arises as a result of compactification of pure continuous Yang-Mills theory from $\lr{d+1}$ dimensions down to $\lr{0+1}$ dimensions. The coupling constant $g_{YM}$ in the compactified theory has the dimension of ${\rm mass}^3$. In the Hamiltonian (\ref{matrix_mechanics_hamiltonian}), we have implicitly set this coupling to unity by expressing all dimensionful quantities in units of $g_{YM}$. Thus the strong-coupling regime of the Hamiltonian (\ref{matrix_mechanics_hamiltonian}) corresponds to energy scales or temperatures smaller than $g_{YM} = 1$, and the weak-coupling regime corresponds to high energies/temperatures.

The dynamics described by the Hamiltonian (\ref{matrix_mechanics_hamiltonian}) is known to be classically chaotic, so that the distance between initially very close points in phase space grows exponentially with time \cite{Savvidy:NPB84}. The fact that such chaotic classical systems effectively forget about initial conditions after some ``thermalization'' time can be interpreted as the formation of a black hole state \cite{Nastase:hep-th/0501068,Berenstein:1104.5469,Berenstein:1211.3425,Hanada:1503.05562} in the framework of holographic duality between compactified super-Yang-Mills theory and the gravitationally interacting system of D0 branes \cite{Witten:96:1}. \ifarxiv{Both black holes and matrix quantum mechanics (\ref{matrix_mechanics_hamiltonian}) are conjectured to be the fastest possible ``scramblers'' of information, with scrambling time of order $O\lr{\log N}$ for $N$ bosonic degrees of freedom \cite{Susskind:08:1}.}

However, most of the previous studies of the real-time dynamics of Yang-Mills-type Hamiltonians similar to (\ref{matrix_mechanics_hamiltonian}) were based on the classical mechanics approximation, which is only justifiable at sufficiently high temperatures. In this work we use the Gaussian state approximation of \cite{Broeckhove:THEOCHEM199} to understand how quantum effects might affect the classically chaotic dynamics of the Hamiltonian (\ref{matrix_mechanics_hamiltonian}). \ifarxiv{We point out that quantum effects tend to decrease the Lyapunov exponents and even make them vanish below some temperature, which might be related to the confinement-deconfinement phase transition \cite{Nishimura:0706.3517}. We argue that such behavior avoids the violation of the Maldacena-Shenker-Stanford (MSS) upper bound $\lambda_L < 2 \pi T$ \cite{Maldacena:1503.01409} on Lyapunov exponents. For classical dynamics the Lyapunov exponents scale with temperature as $\lambda_L \sim T^{1/4}$ \cite{Hanada:1512.00019}, which inevitably violates the MSS bound at sufficiently low temperatures.}

\section{Gaussian state approximation: a simple example with classical chaos}
\label{sec:x2y2}

Before studying the quantum dynamics for the Hamiltonian (\ref{matrix_mechanics_hamiltonian}), in this Section we illustrate the Gaussian state approximation on the example of a simple Hamiltonian
\begin{eqnarray}
\label{x2y2_hamiltonian}
 H = \frac{p_x^2}{2} + \frac{p_y^2}{2} + \frac{x^2 \, y^2}{2} .
\end{eqnarray}
This is one of the simplest reductions of the Yang-Mills-type Hamiltonian (\ref{matrix_mechanics_hamiltonian}) which still features classical chaos \cite{Savvidy:NPB84}.

We start our derivation of quantum corrections to the classical dynamics with the Hamiltonian (\ref{x2y2_hamiltonian}) from the Heisenberg equations for the canonically conjugate operators $\hat{x}$, $\hat{y}$ and $\hat{p}_x$, $\hat{p}_y$:
\begin{eqnarray}
\label{x2y2_heisenberg}
 \partial_t \hat{x}   = \hat{p}_x, \quad
 \partial_t \hat{y}   = \hat{p}_y, \quad
 \partial_t \hat{p}_x = - \hat{x} \, \hat{y}^2, \quad
 \partial_t \hat{p}_y = - \hat{y} \, \hat{x}^2 .
\end{eqnarray}
We can obtain the equations of motion for the expectation values $\vev{\hat{p}_x}$, $\vev{\hat{p}_y}$, $\vev{\hat{x}}$, $\vev{\hat{y}}$ by averaging equations (\ref{x2y2_heisenberg}) with respect to some density matrix $\hat{\rho}$. These equations will contain expectation values of the form $\vev{ \hat{x} \, \hat{y}^2 }$, which should be evolved according to yet another equation, including in turn expectation values of yet larger number of coordinate/momentum operators.

Without any simplifying assumptions we thus get an infinite hierarchy of equations, for which any practical solution is hardly possible. To truncate this infinite set of equations, we approximate the density matrix $\hat{\rho}$ by the most general time-dependent Gaussian function, so that expectation values of products of multiple coordinate and momentum operators can be expressed in terms of the coordinates of the wave packet centers and wave packet dispersion using the Wick's theorem\ifarxiv{\footnote{By Wick's theorem here we mean the well-known general prescription for the integrals of  multi-variable polynomials with the Gaussian weight. In statistics, the same formulae are known as the Isserlis' theorem.}}. It is actually more convenient to characterize the most general Gaussian state by the Wigner function
\begin{eqnarray}
\label{gaussian_wigner_func}
 \rho\lr{\xi} = \mathcal{N} \expa{- \frac{1}{2} \, \lr{\xi - \bar{\xi}} \Sigma^{-1} \lr{\xi - \bar{\xi}} } ,
\end{eqnarray}
where $\xi = \lrc{x, y, p_x, p_y}$ is the four-component vector of the classical phase space variables, the parameters $\bar{\xi}_a \equiv \vev{\hat{\xi}_a}$ describe the center of the Gaussian wave packet and the matrix
\begin{eqnarray}
\label{sigma_matrix}
 \Sigma_{ab}
 =
 \cev{\xi_a \xi_b}
 \equiv
 \vev{\frac{\hat{\xi}_a \hat{\xi}_b + \hat{\xi}_b \hat{\xi}_a}{2} }
 -
 \vev{\hat{\xi}_a} \, \vev{\hat{\xi}_b} .
\end{eqnarray}
characterizes the dispersion of the wave packet in both coordinate and momentum space.

We can now average the Heisenberg equations (\ref{x2y2_heisenberg}) over our Gaussian state with the Wigner function (\ref{gaussian_wigner_func}). Using Wick's theorem, we obtain
\begin{eqnarray}
\label{x2y2_vev_eqs}
 \partial_t x   = p_x, \quad
 \partial_t y   = p_y, \quad
 \partial_t p_x = - x \, \vev{y^2} - 2 \cev{x y} y , \quad
 \partial_t p_y = - y \, \vev{x^2} - 2 \cev{x y} x ,
\end{eqnarray}
where $x \equiv \vev{\hat{x}}$, $y \equiv \vev{\hat{y}}$, $p_x \equiv \vev{\hat{p}_x}$, $p_y \equiv \vev{\hat{p}_y}$. In order to obtain the evolution equations for the dispersions $\cev{\xi_a \xi_b}$, we again use the quantum Heisenberg equations of motion to express the time derivatives of the operator products $\hat{\xi}_a \hat{\xi}_b$ as
\begin{eqnarray}
\label{x2y2_cev_eqs1}
 \partial_t \lr{ \hat{x} \hat{x} }    = \hat{p}_x \hat{x} + \hat{x} \hat{p}_x,
 \quad
 \partial_t \lr{ \hat{x} \hat{y} }    = \hat{p}_x \hat{y} + \hat{x} \hat{p}_y, 
 \quad
 \partial_t \lr{ \hat{x} \hat{p}_x }  = \hat{p}_x^2  -  \hat{x}^2 \hat{y}^2,
 \nonumber \\
 \partial_t \lr{ \hat{x} \hat{p}_y }  = \hat{p}_x \hat{p}_y  -  \hat{x}^3 \hat{y}, 
 \quad
 \partial_t \lr{ \hat{p}_x \hat{p}_x} = - \hat{x} \hat{y}^2 \hat{p}_x - \hat{p}_x \hat{x} \hat{y}^2,
 \quad
 \partial_t \lr{ \hat{p}_x \hat{p}_y}  = - \hat{x} \hat{y}^2 \hat{p}_y - \hat{p}_x \hat{y} \hat{x}^2 .
\end{eqnarray}
Similar equations for other combinations of $\hat{x}$, $\hat{y}$ and $\hat{p}_x$, $\hat{p}_y$ can be obtained by straightforward permutations $\hat{x} \leftrightarrow \hat{y}$, $\hat{p}_x \leftrightarrow \hat{p}_y$. Averaging the equations (\ref{x2y2_cev_eqs1}) over our Gaussian state with the Wigner function (\ref{gaussian_wigner_func}), we obtain the evolution equations for expectation values $\vev{\hat{\xi}_a \hat{\xi}_b}$ of two canonical operators, and, after subtracting the disconnected contributions $\vev{\hat{\xi}_a} \vev{\hat{\xi}_b}$, for the dispersions $\cev{\hat{\xi}_a \hat{\xi}_b}$. The only detail which one has to remember is that the ``classical'' expectation values $\int \mathcal{D}\xi \rho\lr{\xi} \, \xi_a \xi_b \xi_c \ldots $ with the Wigner function $\rho\lr{\xi}$ correspond to quantum expectation values $\tr\lr{\hat{\rho} \lrc{\hat{\xi}_a \hat{\xi}_b \hat{\xi}_c \ldots}_s}$ of the symmetrized operator products, which can be recursively defined as
\begin{eqnarray}
\label{operator_symmetrization}
 \lrc{ \hat{O}\lr{\hat{x}_i} \hat{p}_{i_1} \ldots \hat{p}_{i_n} }_s
 =
 \frac{1}{2} \hat{p}_{i_1} \lrc{ \hat{O}\lr{\hat{x}_i} \hat{p}_{i_2} \ldots \hat{p}_{i_n} }_s
 +
 \frac{1}{2} \lrc{ \hat{O}\lr{\hat{x}_i}  \hat{p}_{i_2} \ldots \hat{p}_{i_n} }_s \hat{p}_{i_1} \iftoggle{arxiv}{,}{.}
\end{eqnarray}
\ifarxiv{where the indices $i$, $i_1, \ldots, i_n$ label the coordinates $x, y$ and $\hat{O}\lr{\hat{x}_i}$ is an arbitrary operator constructed from coordinate operators $\hat{x}$, $\hat{y}$ only. Operator products can be easily transformed into symmetrized form using the canonical commutation relations. After that one can express the expectation values as integrals over the classical phase space $\xi_a$, and use Wick theorem to perform the integrations.} Applying this procedure to \ifarxiv{both left- and right-hand sides of} equations (\ref{x2y2_cev_eqs1}), we obtain \ifarxiv{the following evolution equations for dispersions $\cev{\xi_a \xi_b}$}:
\begin{eqnarray}
\label{x2y2_cev_eqs}
 \partial_t \cev{ x^2 }    = 2 \cev{p_x x},
 \quad
 \partial_t \cev{ x y }    = \cev{p_x y} + \cev{x p_y}, 
 \nonumber \\
 \partial_t \cev{ x p_x }  = \cev{p_x p_x} - \cev{x^2} \vev{y^2} - 2 \cev{x y} \vev{x y},
 \nonumber \\
 \partial_t \cev{ x p_y }  = \cev{p_x p_y}  - \cev{x y} \vev{x^2} - 2 \cev{x^2} \vev{x y}, 
 \nonumber \\
 \partial_t \cev{ p_x p_x} = - 2 \cev{x p_x} \vev{y^2} - 4 \cev{y p_x} \vev{x y},
 \nonumber \\
 \partial_t \cev{ p_x p_y}  = - \cev{x p_y} \vev{y^2} - 2 \cev{y p_y} \vev{x y}
                              - \cev{y p_x} \vev{x^2} - 2 \cev{x p_x} \vev{x y} .
\end{eqnarray}
Equations with other combinations of $x$, $y$ and $p_x$, $p_y$ can be obtained by interchanging $x \leftrightarrow y$.

The equations (\ref{x2y2_vev_eqs}) and (\ref{x2y2_cev_eqs}) form the complete system of equations for the time evolution of the coordinates $\bar{\xi}_a = \lrc{x, y, p_x, p_y}$ of the wave packet center and the dispersion matrix $\cev{\xi_a \xi_b}$. In contrast to the full Schr\"{o}dinger equation, the number of variables in the equations (\ref{x2y2_vev_eqs}) and (\ref{x2y2_cev_eqs}) grows only quadratically in the number of degrees of freedom\ifarxiv{, which makes it particularly suitable for studying complex many-body systems}.

\ifarxiv{An important property of the evolution equations (\ref{x2y2_vev_eqs}) and (\ref{x2y2_cev_eqs}) is that they evolve pure Gaussian states into pure Gaussian states, and, more generally, conserve the von Neumann entropy $S = \tr\lr{\hat{\rho} \, \log \hat{\rho}}$. For the general many-body Gaussian states the latter can be expressed in terms of the symplectic eigenvalues of the correlation matrix $\Sigma_{a b} = \cev{\xi_a \xi_b}$. Symplectic eigenvalues of $\Sigma_{a b}$ are related to the ordinary eigenvalues of the matrix $\epsilon_{a c} \Sigma_{c b}$, where $\epsilon_{a c}$ is the standard symplectic form for the canonical set of coordinates $\xi_a$. One can demonstrate that the equations (\ref{x2y2_vev_eqs}) and (\ref{x2y2_cev_eqs}) conserve the symplectic eigenvalues of $\Sigma_{a b}$. Furthermore, for pure Gaussian states describing $N$ degrees of freedom, the number of independent variables describing the $\lr{2 N} \times \lr{2 N}$ symmetric real matrix $\Sigma_{ab}$ can be reduced to two $\lr{N \times N}$ symmetric real matrices, with entries which can be interpreted as canonically conjugate pairs of coordinates and momenta. This allows to rewrite the equations (\ref{x2y2_vev_eqs}) and (\ref{x2y2_cev_eqs}) as classical equations of motion which follow from a certain extension of the classical Hamiltonian \cite{Broeckhove:THEOCHEM199}. In particular, this allows to devise stable leapfrog-type numerical integrators for (\ref{x2y2_vev_eqs}) and (\ref{x2y2_cev_eqs}). Furthermore, just as the full Schr\"{o}dinger equation $\partial_t \ket{\psi} = i \hat{H} \ket{\psi}$ can be derived by extremizing the ``quantum'' action $S_q = \int dt \bra{\psi} \lr{\partial_t - i \hat{H}} \ket{\psi}$ with respect to all possible time histories of $\ket{\psi}$, the equations (\ref{x2y2_vev_eqs}) and (\ref{x2y2_cev_eqs}) can be obtained by restricting this extremization to the space of all possible time-dependent pure Gaussian states \cite{Broeckhove:THEOCHEM199}.}

\ifarxiv{\section{Comparison with the exact solution of the Schr\"{o}dinger equation}
\label{sec:x2y2_test}}

Let us now compare the numerical solution of the equations (\ref{x2y2_vev_eqs}) and (\ref{x2y2_cev_eqs}) with the solution of the full Schr\"{o}dinger equation. We consider the pure Gaussian state with initially nonzero expectation values of $x$ and $y$ and the minimal quantum dispersion saturating the uncertainty relation:
\begin{eqnarray}
\label{x2y2_initial}
 x = 0.625 \, f, \quad y = 0.325 \, f,
 \quad
 \cev{x x} = \cev{y y} = \cev{p_x p_x} = \cev{p_y p_y} = 1/2 ,
 \nonumber \\
 p_x = p_y = \cev{x y} = \cev{p_x p_y} = \cev{p_x x} = \cev{p_x y} = \cev{p_y y} = \cev{p_x x} = 0 .
\end{eqnarray}
The variable $f$ controls the dominance of the ``classical'' expectation values $\vev{x}$ and $\vev{y}$ over the quantum dispersions. \ifarxiv{At large $f$ we are in the strong-field regime in which the classical equations of motion should be valid, at least for some time from the beginning of the evolution.}

To solve the time-dependent Schr\"{o}dinger equation, we replace the continuum coordinates $x$ and $y$ by a finite lattice with spacing $a$, with sites labelled by indices $i, j \in \lrs{-N_s+1, N_s}$: $x_i = a i$, $y_j = a j$. The potential $V\lr{x, y} = x^2 y^2 / 2$ is turned into a periodic function $V_{ij} = \frac{1}{2} \lr{\frac{2 N_s a}{\pi}}^2 \, \sin^2\lr{\frac{\pi x_i}{2 N_s a}} \, \sin^2\lr{\frac{\pi y_j}{2 N_s a}}$ on this discrete periodic lattice. $V_{ij}$ coincides with the continuum $V\lr{x, y}$ for sufficiently small $x_i$, $y_j$. The operator $\hat{p}^2 = -\partial_x^2$ is replaced by the sum of the usual lattice Laplacians $-\Delta_{ii'} = 2 \delta_{i \, i'} - \delta_{i \, i'+1} - \delta_{i \, i' - 1}$ for both $x$ and $y$ coordinates. We then perform the leapfrog-type evolution by multiplying the wave function $\psi\lr{x, y} \rightarrow \psi_{ij}$ by the unitary evolution operator
\begin{eqnarray}
\label{leapfrog_evolution_operator}
 U\lr{\delta t} = \expa{i V {\delta t}/2} \expa{i \frac{\hat{p}^2}{2} {\Delta t}} \expa{i V {\delta t}/2}
\end{eqnarray}
with sufficiently small time step ${\delta t}$. \ifarxiv{This procedure is numerically stable, has discretization error of order of ${\delta t}^2$ and preserves the norm of the discretized wave function.} In order to control discretization and finite-volume artifacts, in addition to the solution with lattice spacing $a$, time step ${\delta t}$ and lattice size $N_s$ we also consider solutions with parameters $\lrc{2 a, {\delta t}, N_s}$, $\lrc{a, 2 {\delta t}, N_s}$ and $\lrc{a, {\delta t}, N_s/2}$. We then estimate the discretization and finite volume error of the expectation values $\vev{\hat{O}\lr{t}}$ as the difference between their minimal and maximal values over simulations with these four sets of parameters.

\begin{figure}[h!tb]
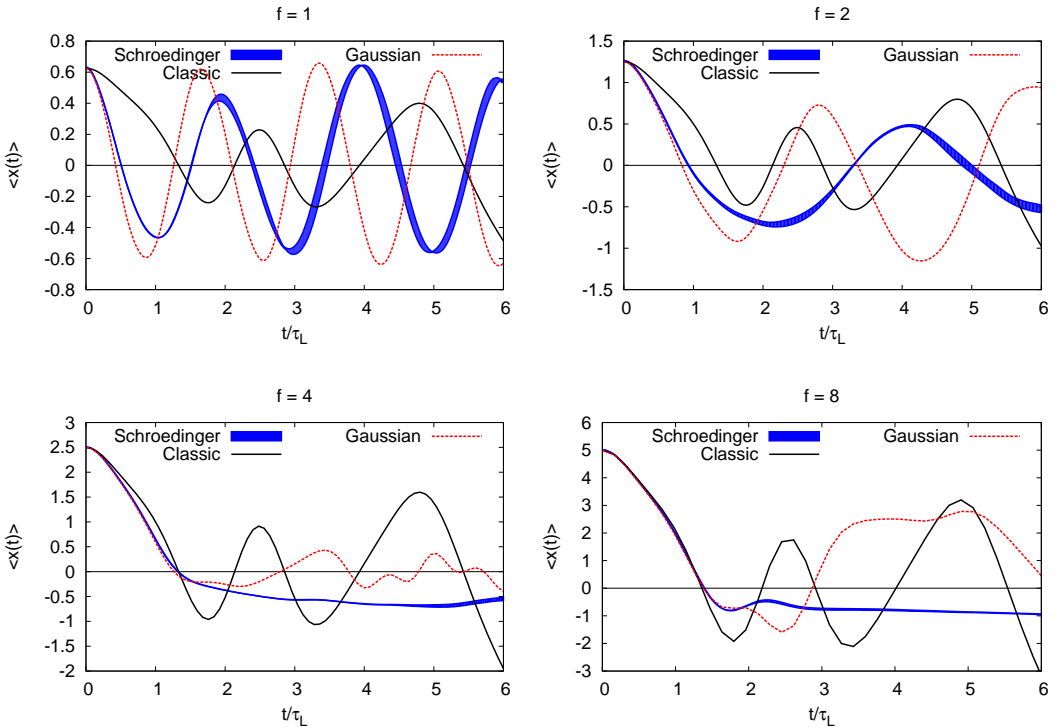

  \centering
  \includegraphics[width=0.35\textwidth,angle=-90]{{{test_f1}}}\includegraphics[width=0.35\textwidth,angle=-90]{{{test_f2}}}\\
  \includegraphics[width=0.35\textwidth,angle=-90]{{{test_f4}}}\includegraphics[width=0.35\textwidth,angle=-90]{{{test_f8}}}\\
  \caption{Time dependence of the expectation values $\vev{x \lr{t}}$, obtained from the classical equations of motion (solid black line), from the Gaussian state approximation (dotted red line) and from the numerical solution of the Schr\"{o}dinger equation (blue strip with strip width estimating the finite-spacing and finite-volume artifacts). The initial state is the Gaussian state with center and dispersions given by (\ref{x2y2_initial}), with $f = 1$ (top left), $f = 2$ (top right), $f = 4$ (bottom left) and $f = 8$ (bottom right). The time $t$ is in units of the inverse classical Lyapunov time $\tau_L^c \equiv \lr{\lambda_L^c}^{-1} \sim f^{-1}$.}
  \label{fig:x2y2_comparison}
\end{figure}

In Fig.~\ref{fig:x2y2_comparison} we compare the time dependence of the expectation value $\vev{\hat{x}\lr{t}}$ of the $x$ coordinate, obtained from the classical equations of motion, from the Gaussian state approximation, and from the numerical solution of the Schr\"{o}dinger equation. From the topmost plot on the left to the lowest plot on the right we increase the initial expectation values of $x$ and $y$ coordinates (factor $f$ in (\ref{x2y2_initial})) as compared to the quantum dispersions which are fixed for all $f$. We thus move from the quantum regime at small $f$ to the classical regime at large $f$. Since the Hamiltonian (\ref{matrix_mechanics_hamiltonian}) is a reduction of the Hamiltonian (\ref{matrix_mechanics_hamiltonian}), we can again interpret the small-$f$ classical regime as the strong-coupling regime, and the large-$f$ classical regime as the weak-coupling regime. We note that as $f$ becomes larger, the classical solution is rescaled as $x\lr{t} \rightarrow f x\lr{f t}$ - that is, the classical dynamics becomes faster, and the corresponding classical Lyapunov exponent $\lambda_L^c$ becomes $f$ times larger. To make the comparison of different real-time evolution methods independent of this trivial classical scaling of the Lyapunov exponent, we express the physical time $t$ in units of the classical Lyapunov time $\tau_L^c \equiv \lr{\lambda_L^c}^{-1}$.

One can see that when the expectation values $\vev{x}$, $\vev{y}$ are not very large compared to the corresponding quantum dispersions ($f = 1$, $f = 2$), the difference between the classical solution and the solution of the Schr\"{o}dinger equation is rather large even at early evolution times. The Gaussian state approximation is much closer to the full quantum evolution for $\lambda_L^c \, t \lesssim 1$. For $f = 1$, the Gaussian state approximation also captures the period and the amplitude of the oscillations of $\vev{\hat{x}\lr{t}}$ rather well. At larger $f = 4$ and $f = 8$ we see how the classical dynamics becomes more and more exact at early times $\lambda_L^c \, t \lesssim 1$, and the numerical accuracy of the Gaussian state approximation extends even further to $\lambda_L^c \, t \lesssim 2$. In contrast, the late-time dynamics is captured less and less precisely as the initial expectation values $\vev{x}$, $\vev{y}$ become larger. In particular, while at $f = 4$ the Gaussian state approximation describes rather well the suppression of oscillations of $\vev{\hat{x}\lr{t}}$ as compared to the classical dynamics, for the largest $f = 8$ at late times both the classical dynamics and the Gaussian state approximation are equally inaccurate.

\ifarxiv{To summarize, it seems that in the quantum (strong-coupling) regime with comparable values of coordinate expectation values and their dispersions the Gaussian approximation (\ref{x2y2_vev_eqs}), (\ref{x2y2_cev_eqs}) is qualitatively good at both early and late times, but is not precise at the quantitative level. In the classical (weak-coupling) regime, with expectation values dominating over quantum dispersions, the Gaussian approximation becomes quantitatively precise at early times, but qualitatively wrong at late times. Probably the reason for this behavior is that the classically unstable early-time dynamics of the wave packet center at large $f$ leads to the fast spreading of the wave function which becomes strongly non-Gaussian. We also note that the expectation value $\vev{\hat{x}\lr{t}}$ at late times $t$ also approaches zero for the full Schr\"{o}dinger equation, although at time scales much longer than the range of the plots in Fig.~\ref{fig:x2y2_comparison}. This ``slowing down'' of the full quantum dynamics can be roughly understood as follows. The relevant time scale at late times is set by the production rate $\partial_t S_{KS} \sim \lambda_L \sim f$ of the Kolmogorov-Sinai entropy per unit of phase space volume $V$, which grows as $V \sim f^2$. Hence the ratio $\partial_t S_{KS}/V$ decreases with the amplitude $f$ as $f^{-1}$.}

Finally, an important methodological question is how to control the validity of the Gaussian state approximation without having the reference solution of the Schr\"{o}dinger equation, which is practically unfeasible for sufficiently large number of degrees of freedom. One of the possible intuitive criteria is the importance of terms which contain squares of quantum dispersions in the potential energy $\vev{V\lr{x, y}} = \vev{x^2 y^2}/2 = x^2 y^2 + x^2 \cev{y^2} + y^2 \cev{x^2} + 4 x y \cev{x y} + 2 \cev{x y}^2 + \cev{x^2} \cev{y^2}$. We have found that the Gaussian state approximation starts being inaccurate when the last two summands in the above expression for $\vev{V\lr{x, y}}$ become comparable with the kinetic energy $\vev{p_x^2}/2 + \vev{p_y^2}/2$. This replaces the condition of large occupation numbers for the classical dynamics.

\section{Gaussian state approximation for matrix quantum mechanics}
\label{sec:bfss_gaussian}

In this Section, we apply the Gaussian state approximation to the matrix quantum mechanics with the Hamiltonian (\ref{matrix_mechanics_hamiltonian}). To this end we decompose the matrix-valued canonical variables $X_i$ and $P_i$ in the basis of generators $T_a$ of $SU\lr{N}$ Lie algebra, with $\tr\lr{T_a T_b} = \delta_{a b}$ and $\lrs{T_a, T_b} = i C_{abc} T_c$: $X_i = X_i^a T_a$, $P_i = P_i^a T_a$. Following the construction of the Gaussian state approximation outlined in Section~\ref{sec:x2y2}, we now average the Heisenberg equations of motion with the Hamiltonian (\ref{matrix_mechanics_hamiltonian}) over the most general Gaussian state characterized by the expectation values $X^a_i = \vev{\hat{X}^a_i}$, $P^a_i\lr{t} = \vev{\hat{P}^a_i}$ and the dispersions $\cev{X^a_i X^b_j}$, $\cev{X^a_i P^b_j}$ and $\cev{P^a_i P^b_j}$ defined as in (\ref{sigma_matrix}). We obtain the following equations of motion:
\begin{eqnarray}
\label{bfss_vev_eqs}
 \partial_t X^a_i = P^a_i,
 \quad
 \partial_t \cev{ X^a_i X^b_j } = \cev{ X^a_i P^b_j } + \cev{ X^b_j P^a_i },
 \quad
 \partial_t P^a_i = - C_{abc} C_{cde} X^b_j X^d_i X^e_j
 - \nonumber \\
 - C_{abc} C_{cde} X^b_j \cev{ X^d_i X^e_j}
 - C_{abc} C_{cde} \cev{X^b_j X^e_j} X^d_i
 - C_{abc} C_{cde} \cev{X^b_j X^d_i} X^e_j ,
\end{eqnarray}
\begin{eqnarray}
 \partial_t \cev{ X^a_i P^f_k } = \cev{ P^a_i P^f_k }
 -
 C_{abc} C_{cde} \vev{X^d_i X^e_j} \cev{ X^b_j X^f_k }
 - \nonumber \\ -
 C_{abc} C_{cde} \vev{X^b_j X^e_j} \cev{ X^d_i X^f_k }
 - 
 C_{abc} C_{cde} \vev{X^b_j X^d_i} \cev{ X^e_j X^f_k } \nonumber  ,
\end{eqnarray}
\begin{eqnarray}
 \partial_t \cev{ P^a_i P^f_k } =
 -
 C_{abc} C_{cde} \vev{X^d_i X^e_j } \cev{ X^b_j P^f_k }
 - \nonumber \\ -
 C_{abc} C_{cde} \vev{X^b_j X^e_j } \cev{ X^d_i P^f_k }
 - 
 C_{abc} C_{cde} \vev{X^b_j X^d_i} \cev{ X^e_j P^f_k }
 + \lr{ \lrc{a,i} \leftrightarrow \lrc{f, k} } \nonumber .
\end{eqnarray}
\ifarxiv{If quantum dispersions can be neglected, these equations reduce to the well-known classical equations of motion
\begin{eqnarray}
\label{bfss_classic_eqs}
 \partial_t X^a_i = P^a_i,
 \quad
 \partial_t P^a_i = - C_{abc} C_{cde} X^b_j X^d_i X^e_j .
\end{eqnarray}}

\ifarxiv{The equations (\ref{bfss_vev_eqs}) can be solved numerically using the Runge-Kutta or leapfrog integrators.The number of Flops per time step scales as $N^5$ and is saturated by the double commutator terms similar to $C_{abc} C_{cde} \vev{X^b_j X^d_i} \cev{ X^e_j P^f_k }$. The memory requirements scale as $N^4$.}

\ifarxiv{
\section{Quantum Lyapunov exponents}
\label{sec:quantum_lyapunov}
}

To study how quantum effects influence the classically chaotic dynamics of the Hamiltonian (\ref{matrix_mechanics_hamiltonian}), we extend the definition of Lyapunov exponents to the quantum case by considering two solutions of the equations (\ref{bfss_vev_eqs}) for which the initial Gaussian states differ by a very small shift of the wave packet center $X^a_i \rightarrow X^a_i + \epsilon^a_i$. For the chaotic system, the difference $\delta X^a_i$ between the expectation values $X^a_i$ for these two solutions should first grow exponentially as $|\delta X^a_i| \sim |\epsilon^a_i| \expa{\lambda_L t}$, where $\lambda_L$ is the leading Lyapunov exponent, and then saturate at a typical scale set by the system size. It is also easy to see that this definition is equivalent to the ``out-of-time-order'' correlator
\begin{eqnarray}
\label{oto_correlator}
 \delta X^a_i = \bra{in} e^{i \epsilon^a_i \hat{P}^a_i } \hat{X}^b_j\lr{t} e^{- i \epsilon^a_i \hat{P}^a_i } \ket{in} - \bra{in} \hat{X}\lr{t} \ket{in} = i \bra{in} \lrs{\hat{P}^a_i\lr{0}, \hat{X}^b_j\lr{t} } \ket{in} \epsilon^a_i .
\end{eqnarray}
\ifarxiv{We put ``out-of-time-order'' in quotes, since in the conventional definition of out-of-time-order-correlators, the commutator $\lrs{\hat{P}^a_i\lr{0}, \hat{X}^b_j\lr{t} }$ is squared inside the expectation value (see e.g. \cite{Maldacena:1503.01409}). The fact that in our numerical setup the expectations $\vev{\hat{X}^a_i\lr{t}}$ of single-coordinate operators are nonzero allows us to relate the expectation value of $\lrs{\hat{P}^a_i\lr{0}, \hat{X}^b_j\lr{t} }$ to Lyapunov exponents without squaring.}

Our simulations were performed with $d = 9$ compactified spatial dimensions (motivated by the BFSS matrix model \cite{Susskind:97:1}) and $SU\lr{5}$ Lie algebra. The initial values of the coordinates $X^a_i\lr{t = 0}$ of the wave packet center were randomly drawn from the Gaussian distribution with dispersion $\sigma^2$, the initial values of momenta $P^a_i\lr{t = 0}$ were set to zero. The quantum dispersions were set to the values $\cev{X^a_i X^b_j} = \delta^{ab} \delta_{ij} \, \sigma_{xx}$, $\cev{P^a_i P^b_j} = \delta^{ab} \delta_{ij} \, \sigma_{pp}$, $\cev{X^a_i P^b_j} = 0$, $\sigma_{xx} = \lr{N \lr{d - 1}}^{-1/3}/2$, $\sigma_{pp} = 1/\lr{4 \sigma_{xx}}$ which minimize the expectation value of the Hamiltonian (\ref{matrix_mechanics_hamiltonian}) in the space of pure Gaussian states\ifarxiv{{ }and thus mimick the ground state within the Gaussian state approximation}. We choose $\epsilon^a_i$ to be a random vector with $|\epsilon^a_i| = 0.00001$.

\begin{figure}
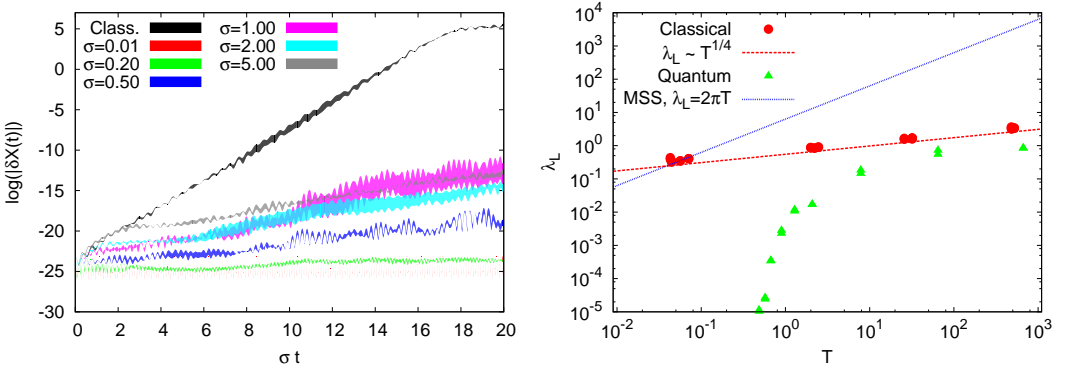

  \centering
  \includegraphics[angle=-90,width=0.5\textwidth]{{{lyapunov}}}\includegraphics[angle=-90,width=0.5\textwidth]{{{mss_bound_bfss}}}\\
  \caption{A comparison of Lyapunov instability of the matrix quantum mechanics with the Hamiltonian (\ref{matrix_mechanics_hamiltonian}) in the classical and in the Gaussian state approximations. On the left: time dependence of the Lyapunov distance between two initially close solutions of the evolution equations. On the right: temperature dependence of the leading Lyapunov exponents for classical dynamics (eqs.~(\ref{bfss_classic_eqs})) and quantum dynamics (eqs.~(\ref{bfss_vev_eqs})) compared with the MSS bound. We change the classical dispersion $\sigma$ of the initial values of $X^a_i$ coordinates and keep fixed the initial quantum dispersion. On the right plot, the equations of state (\ref{classical_eos}) and (\ref{quantum_eos}) are used to express temperature in terms of energy.}
  \label{fig:lyapunov}
\end{figure}

We illustrate the time dependence of the distance $|\delta X^a_i|$ between two close solutions of the equations (\ref{bfss_vev_eqs}) in the left plot on Fig.~\ref{fig:lyapunov}. For comparison we also show the time dependence of $|\delta X^a_i|$ for the classical equations of motion. Since even the classical Lyapunov exponents depend on the scale of $X^a_i$ variables as $\lambda_L^c \sim \sigma$, we use $\sigma \, t \sim \lambda_L^c \, t$ for the time coordinate. The classical dynamics exhibits a clear exponential growth $|\delta X\lr{t}| \sim \expa{\lambda_L \, t}$ with time, followed by the expectable saturation of Lyapunov distance at late times for a bound system. For quantum dynamics the growth rate of the Lyapunov distance coincides with the classical one only at some initial period of time, which becomes larger as the dynamics becomes more classical at larger $\sigma$. Presumably, this initial period of time can be related to the thermalization of our ``artificial'' initial state itself, which is then followed by the decay of small perturbations on top of the more universal thermalized state. At later times the leading Lyapunov exponent $\lambda_L$ for the quantum dynamics turn out to be several times smaller. \ifarxiv{Looking at the plots on Fig.~\ref{fig:x2y2_comparison}, this finding does not seem surprising, as it seems that in the quantum regime the classically chaotic system would tend to forget about initial conditions by spreading its wave function, rather than exhibiting complex oscillations of the coordinate/momenta expectation values.}

Let us now consider the temperature dependence of the quantum Lyapunov exponents obtained numerically from the correlator (\ref{oto_correlator}). In order to introduce temperature, we rely on the chaotic, ``self-averaging'' classical dynamics of the Hamiltonian (\ref{matrix_mechanics_hamiltonian}) which implies the equivalence between canonical and micro-canonical ensembles. At sufficiently high energies and temperatures we can use the classical equation of state \cite{Nishimura:0710.2188,Hanada:1512.00019}
\begin{eqnarray}
\label{classical_eos}
 E = \chi \, T, \quad \chi = \frac{3}{4} \, \lr{\lr{d-1} \lr{N^2 - 1} - \frac{d \lr{d - 1}}{2}}
\end{eqnarray}
to translate energy $E$ into temperature $T$. At low temperatures, this equation of state changes due to quantum effects. In particular, in the limit of zero temperature the energy approaches some finite value $E_0$ of the ground-state energy \cite{Nishimura:0706.3517}. To be consistent with the Gaussian state approximation, the ground state energy should be defined by minimizing the Hamiltonian (\ref{matrix_mechanics_hamiltonian}) over all pure Gaussian states, which yields $E_0 = \frac{3}{8} d \lr{d - 1}^{1/3} N^{1/3} \lr{N^2 - 1}$. In order to interpolate between the low-temperature asymptotics $E \rightarrow E_0$ (at $T \rightarrow 0$) and the classical high-temperature equation of state (\ref{classical_eos}), we use a phenomenological formula
\begin{eqnarray}
\label{quantum_eos}
 E = E_0/\tanh\lr{\frac{E_0}{\chi \, T}}, \quad
 E_0 = \frac{3}{8} d \lr{d - 1}^{1/3} N^{1/3} \lr{N^2 - 1} .
\end{eqnarray}
Since the first-principle equation of state of \cite{Nishimura:0706.3517} would be anyway inconsistent with the Gaussian state approximation, we prefer to use the artifical equation of state (\ref{quantum_eos}). The most important conclusions of this work should not dramatically depend on moderate changes in the equation of state.

Temperature dependence of quantum Lyapunov exponents is illustrated on the right plot on Fig.~\ref{fig:lyapunov}. For comparison, we also show the temperature dependence of the classical Lyapunov exponents, which is well described by the formula $\lambda_L = \lr{0.292 - 0.42/N^2} T^{1/4}$ \cite{Hanada:1512.00019}, with the temperature $T$ defined from (\ref{classical_eos}). We see that quantum corrections tend to decrease the Lyapunov exponents at intermediate temperatures. Moreover, at some ``critical'' temperature $T_c \approx 0.6$ the leading Lyapunov exponent seem to vanish. Interestingly, first-principle Monte-Carlo simulations indicate that in the large-$N$ limit the matrix quantum mechanics (\ref{matrix_mechanics_hamiltonian}) has a finite-temperature confinement-deconfinement phase transition at roughly the same temperature $T_c' \approx 0.9$ \cite{Nishimura:0706.3517}.  This has to be contrasted with the supersymmetric BFSS model, which is not confining all the way down to the strong-coupling regime at $T = 0$\ifarxiv{{ }\cite{Hanada:0707.4454}  (let us remind that for the Hamiltonian (\ref{matrix_mechanics_hamiltonian}), strong-coupling regime corresponds to low temperatures)}. By analogy with $\mathcal{N}=4$ super-Yang-Mills in $D = 3+1$, we expect the MSS bound $\lambda_L < 2 \pi T$ \cite{Maldacena:1503.01409} to be saturated in this regime. On the other hand, in the confinement phase of the non-supersymmetric Hamiltonian (\ref{matrix_mechanics_hamiltonian}), it is natural to expect that Lyapunov exponents are $1/N$-suppressed, because the large-N limit is temperature independent \cite{Nishimura:0706.3517}
and hence is equivalent to zero temperature. Our numerical results shown on the right plot on Fig.~\ref{fig:lyapunov} thus suggest that the growth of quantum corrections at low temperatures and probably the confinement-deconfinement transition prevent the violation of MSS bound, which for classical Lyapunov exponents happens at the temperature $T^{\star} = \lr{\frac{0.292 - 0.42/N^2}{2 \pi}}^{4/3} = 0.015$ which is much lower than $T_c$.

\section{Discussion and outlook}
\label{sec:conclusions}

In these Proceedings we have outlined the Gaussian state approximation for the real-time dynamics of many-body systems, used previously in the quantum chemistry context \cite{Broeckhove:THEOCHEM199}, and demonstrated on a simple example that it is capable of reproducing the essential features of quantum dynamics which are absent for the classical equations of motion. We applied this approximation to study the real-time dynamics of the matrix quantum mechanics, and found that quantum corrections tend to decrease the leading Lyapunov exponent $\lambda_L$ extracted from the simplest out-of-time-order correlator $\bra{in} \lrs{\hat{P}^a_i\lr{0}, \hat{X}^a_i\lr{t}} \ket{in}$. We have shown that quantum corrections seem to be large enough to ensure the validity of the MSS bound $\lambda_L < 2 \pi T$ at sufficiently low temperatures. Of course, we can make this statement only up to the unquantified systematic error of the Gaussian state approximation, which becomes quantitatively worse at low temperatures (see Fig.~\ref{fig:x2y2_comparison} for illustration). In particular, it would be interesting to understand in more details whether the vanishing of quantum Lyapunov exponents below some finite temperature is a signature of the finite-temperature phase transition found in \cite{Nishimura:0706.3517}, or rather an artifact of the Gaussian state approximation.

\ifarxiv{It would be interesting to make contact between our definition (\ref{oto_correlator}) of the Lyapunov distance with the four-point out-of-time-order correlators of the form $\bra{in} \lrs{\hat{P}^a_i\lr{0}, \hat{X}^a_i\lr{t}}^2 \ket{in} \sim \sum_s \bra{in} \lrs{\hat{P}^a_i\lr{0}, \hat{X}^a_i\lr{t}} \ket{s} \bra{s} \lrs{\hat{P}^a_i\lr{0}, \hat{X}^a_i\lr{t}} \ket{in}$, which are more conventionally used in the literature \cite{Maldacena:1503.01409}. We believe that at least for systems with sufficiently large number of degrees of freedom, or sufficiently close to the classical regime, the sum over intermediate states $\ket{s}$ is saturated by the saddle point at $\ket{s} = \ket{in}$ if the correlators $\bra{in} \lrs{\hat{P}^a_i\lr{0}, \hat{X}^a_i\lr{t}} \ket{in}$ are sufficiently large, so that $\bra{in} \lrs{\hat{P}^a_i\lr{0}, \hat{X}^a_i\lr{t}}^2 \ket{in} \sim \bra{in} \lrs{\hat{P}^a_i\lr{0}, \hat{X}^a_i\lr{t}} \ket{in}^2$.}

\ifarxiv{As a further direction of work, one can also study the evolution of quantum entanglement within the Gaussian state approximation, which might provide an alternative definition of the time scale at which quantum systems forget about initial conditions and ``scramble'' information \cite{Susskind:08:1}. In the Gaussian state approximation one can easily characterize the reduced density matrix of a subsystem by restricting the two-point correlators $\cev{\hat{\xi}_a \hat{\xi}_b}$ (see equation (\ref{sigma_matrix})) of canonical variables to the degrees of freedom of this subsystem (see \cite{Berges:1707.05338,Bianchi:1709.00427} for some recent work in this direction).}

The generalization of the Gaussian state approximation to Yang-Mills theory should not meet significant conceptual difficulties. The resulting algorithm would require the number of Flops proportional to the square of the number of (spatial) lattice sites, which should allow for simulations on not very large lattices with modest computational resources.


\iftoggle{arxiv}{

}
{
 \bibliography{Buividovich}
}

\end{document}
